# Bilevel Model for Electricity Market Mechanism Optimisation via Quantum Computing Enhanced Reinforcement Learning

Shuyang Zhu, *Student Member, IEEE*, Ziqing Zhu, *Member, IEEE*

*Abstract*— In response to the increasing complexity of electricity markets due to low-carbon requirements and the integration of sustainable energy sources, this paper proposes a dynamic quantum computing enhanced bilevel optimization model for electricity market operations. The upper level focuses on market mechanism optimization using Reinforcement Learning (RL), specifically Proximal Policy Optimization (PPO), while the lower level models the bidding strategies of Generating Companies (GENCOs) using a Multi-Agent Deep Q-Network (MADQN) enhanced with quantum computing through a Variational Quantum Circuit (VQC). The three main contributions of this work are: (1) establishing a dynamic bilevel model with timely feedback between the upper and lower levels; (2) parameterizing and optimizing market mechanisms to derive the most effective designs; and (3) introducing quantum computing into the context of electricity markets to more realistically simulate market operations. The proposed model is tested on the IEEE 30-bus system with six GENCOs, demonstrating its effectiveness in capturing the complexities of modern electricity markets.

## I. INTRODUCTION

The global imperative to reduce carbon emissions has accelerated the development and integration of sustainable energy sources into power systems worldwide. As renewable energy generation—particularly from wind and solar—continues to increase, the electricity market's dynamics have become increasingly complex [1]. The inherent variability and uncertainty associated with renewable energy sources significantly influence the bidding behaviours of Generating Companies (GENCOs) in day-ahead and real-time markets [2]. Traditional fossil fuel-based GENCOs and renewable energy providers now operate within a highly volatile market environment, necessitating more sophisticated strategies to remain competitive.

In this evolving landscape, market regulators face the critical challenge of designing appropriate market mechanisms that not only accommodate the unique characteristics of renewable energy but also ensure market efficiency, reliability, and fairness. Existing models for GENCO bidding strategies often fall short in capturing the intricate interactions between market participants, especially under high renewable penetration scenarios. These models typically assume static or overly simplified market conditions [2], failing to account for the dynamic and stochastic nature of modern electricity markets. Similarly, current approaches to electricity market mechanism design are limited in their ability to adapt to the rapid changes brought about by the integration of sustainable energy sources [3]. Many models do not adequately parameterize market mechanisms, leading to suboptimal configurations that can hinder the effective participation of renewable energy providers and compromise overall market performance.

To address these shortcomings, we propose a novel bilevel optimization framework that blends traditional reinforcement learning (RL) and an emerging technology, quantum computing (QC), to more accurately simulates electricity market operations under high renewable energy penetration. Our work is pioneering in that it parameterizes market mechanisms, allowing for the optimization of key parameters to derive the most effective market designs. This approach enables the exploration of optimal configurations for bid caps, settlement rules, and penalties for renewable output deviations, which are critical for enhancing market efficiency and fairness. Firstly, the generator bid cap is essential in preventing market manipulation and excessive pricing. By limiting the maximum price that GENCOs can bid, the market operator can prevent individual participants from influencing market prices to their advantage, thereby protecting consumer interests and promoting fair competition. This cap is particularly significant in a market with high renewable energy penetration, as it prevents pricing strategies that could disadvantage less predictable renewable sources like wind and solar. Secondly, the settlement rules, which define how GENCOs are compensated for the electricity they supply, have profound impacts on bidding strategies and market outcomes. The choice between pay-as-bid and pay-as-clear (marginal pricing) mechanisms influences the incentives for GENCOs, potentially affecting market efficiency and the integration of renewable energy. Optimizing settlement rules is crucial to balance transparency, predictability of earnings, and the promotion of competitive bidding behaviours. Thirdly, penalties for deviations in renewable energy output are necessary to maintain grid stability and encourage accurate forecasting by renewable energy providers. However, setting these penalties too high can discourage participation from renewable GENCOs, hindering the progress towards low-carbon energy systems. Therefore, optimizing these penalties is essential to facilitate the integration of renewable energy sources while ensuring the reliability of the power system.

The upper level of our model focuses on market mechanism optimization using RL techniques, specifically employing the Proximal Policy Optimization (PPO) algorithm [4]. By parameterizing the market mechanisms, the upper-level model adjusts these parameters based on feedback from market performance indicators such as social welfare, market concentration indices, and renewable penetration rates. This dynamic adjustment guides the market towards configurations that promote sustainability, competitiveness, and reliability.

The lower level models the bidding strategies of GENCOs using a Multi-Agent Deep Q-Network (MADQN) [5] framework. More importantly, we enhance this layer by integrating quantum computing through the implementation of a Variational Quantum Circuit (VQC) with six qubits—each representing a GENCO. This quantum computing approach allows for a more realistic and sophisticated simulation of the electricity market, capturing the complex, non-linear interactions among market participants. The use of quantum computing in this context is innovative, as it leverages quantum parallelism and entanglement to process vast state and action spaces more efficiently than classical computing methods. By combining advanced RL algorithms with quantum computing techniques in a bilevel framework, our model offers a comprehensive tool for both market participants and regulators. It facilitates the development of optimal bidding strategies for GENCOs while simultaneously enabling regulators to design market mechanisms that are robust to the uncertainties introduced by renewable energy sources. This dual approach ensures that the objectives of maximizing social welfare and promoting sustainable energy integration are met, contributing to the overall stability and efficiency of the electricity market.

The rest of the paper is categorised as follows: Section II introduces the methodology of the proposed bilevel electricity market mechanism optimisation model, followed by a description of the quantum RL algorithm in Section III; Section IV records the case study that experiments with our algorithm and compare it to a classical approach and Section V discusses the main conclusions.

## II. BILEVEL MODEL FRAMEWORK FOR ELECTRICITY MARKET SIMULATION

This section describes the proposed bilevel model framework for the market mechanism optimization problem. The lower level simulates bidding and clearing process of multiple GENCOs in day-ahead electricity market, while the upper level is to optimise the market mechanism based on the simulation results from the lower level model. The section will focus on concluding the design and foundational algorithms for the two levels, as well as detailing how the two levels interact during operation.

### A. Lower Level: GENCOs Bidding Strategy Optimisation

In the real-world electricity market, the trading process at each time-step (which can be half-hourly, hourly, daily, monthly, or annually) typically involves four main steps: Market Information Announcement and call for bids, Bids Submission, Market Clearing and Market Participants' Bidding Strategy Optimisation. The lower level model, referred to as the GENCOs Bidding Strategy Optimization (BSO) Model, focuses on exploring the fourth step by utilizing data from the previous three steps. Analyzing the decision-making processes of Generating Companies (GENCOs) is crucial for market operators to understand and address issues related to market manipulation and reduced social welfare. This analysis aids in designing incentive-compatible market mechanisms that converge to Pareto Optimality, aligning the objectives of market participants seeking higher remunerations with the Independent System Operators' (ISOs) goal of improving total social welfare.

The Markov Decision Process (MDP) offers an intuitive way to model the decision-making processes of GENCOs. In this agent-based modelling approach, each GENCO is considered a "smart agent," and their decision-making process is broken down into states, actions, rewards, and policies that agents follow at each time step. Formally, an MDP is defined as a tuple $\{S, A, R, P\}$. State space ($S$) is the set of all possible states of the environment. In our model, the state $s_t \in S$ at time $t$ represents the total demand in the day-ahead market for each hour. Action space ($A$) refers to the set of all possible actions available to the agents. For each GENCO, the $a_t \in A$ at time $t$ consists of two components: decision on whether to participate in the Bid Submission Stage and if participating, the specific bid within the electricity price cap to submit. This design allows for the possibility of opting out of market participation during certain periods, adding realism to the model. Reward function ($R$) is to map $R: S \times A \times S \to \mathbb{R}$ that assigns a scalar reward $r_t$ to the agent after taking action $a_t$ in state $s_t$ and transitioning to state $s_{t+1}$. The reward for each GENCO considers fixed costs, switching costs, and, for renewable generation participants, penalties for deviations in output. State Transition Probability (P) is a function $P: S \times A \times S \to [0,1]$ defining the probability $P(s_{t+1}|s_t, a_t)$ of transitioning from state $s_t$ to state $s_{t+1}$ after action $a_t$. This captures the dynamics of the electricity market influenced by the actions of all GENCOs. At each time step $t$, each GENCO observes the current state $s_t$, makes a decision $a_t$ based on their policy $\pi$, receives a reward $r_t$, and transitions to the next state $s_{t+1}$. The sequence of states, actions, and rewards forms an episode:

$$\{(s_0, a_0, r_1), (s_1, a_1, r_2), \dots, (s_{T-1}, a_{T-1}, r_T)\} \quad (1)$$

where $T$ is the total number of time steps in the episode (e.g., $T = 24$ for hourly intervals over one day).

Reinforcement Learning (RL) is a type of machine learning where an agent learns to make decisions by performing actions in an environment and receiving feedback in the form of reward. Multi-Agent Reinforcement Learning (MARL) extends this concept to scenarios involving multiple agents, each learning and making decisions independently, which can effectively simulate MDP because they share a high degree of similarity in describing and handling state transitions, reward functions, and policy evolution in multi-agent environments. Therefore, the lower level is to use MARL method to model GENCOs' decision-making process for the day-ahead bidding behaviours in the wholesale market within the horizon of one month. For each day, the time-step is considered as hourly over a 24-hour period. The lower level model uses the Multi-Agent

Deep Q-Network (MADQN) algorithm. Each GENCO agent aims to maximize its expected cumulative reward by learning an optimal $\pi$ based on the estimated quality of actions, represented by the Q-function. The Q-function $Q(s, a; \theta)$ approximates the expected cumulative reward when taking action $a$ in state $s$ and thereafter following policy $\pi$, with $\theta$ being the parameters of the neural network. The Q-learning update rule for agent $i$ at time $t$ is:

$$Q'_i(s_t, a_t^i; \theta_i) \leftarrow Q_i(s_t, a_t^i; \theta_i)$$
$$+ \alpha [r_t^i + \gamma \max_{a'} Q_i(s_{t+1}, a'; \theta_i^-) - Q_i(s_t, a_t^i; \theta_i)] \quad (2)$$

where: : $Q(s_t, a_t^i; \theta_i)$ is the estimated Q-value for the current state $s_t$ and action $a_t^i$ taken by agent $i$, $r$ is the reward received by agent $i$ after taking action $a_t^i$ in state $s_t$, $\gamma$ is the discount factor, $\alpha$ is the learning rate, $s_{t+1}$ is the new state after action $a_t^i$ is taken, and $a'$ represents possible future actions from state $s_{t+1}$, $\theta_i$ and $\theta_i^-$ represent current parameters of agent $i$'s Q-network and parameters of the target network (a delayed copy of $\theta_i$).

The loss function for training the Q-network is:
$$L_i(\theta_i) = \mathbb{E}_{(s_t, a_t^i, r_t^i, s_{t+1})}[(y_t^i - Q_i(s_t, a_t^i; \theta_i))^2] \quad (3)$$
where $\mathbb{E}_{(s_t, a_t^i, r_t^i, s_{t+1})}$ refers to the experience and the target value $y_t^i$ is defined as:
$$y_t^i = r_t^i + \gamma \max_{a'} Q_i(s_{t+1}, a'; \theta_i^-) \quad (4)$$

The gradient of the loss function with respect to the network parameters $\theta_i$ is shown in (5). By minimizing this loss function using stochastic gradient descent, each GENCO agent updates its policy to better approximate the optimal action-value function $Q^*(s, a)$.

$$\nabla_{\theta_i} L_i(\theta_i) = \mathbb{E}_{(s_t, a_t^i, r_t^i, s_{t+1})}[(y_t^i - Q_i(s_t, a_t^i; \theta_i))\nabla_{\theta_i} Q_i(s_t, a_t^i; \theta_i)] \quad (5)$$

To stabilize training and improve sample efficiency, the agents use experience replay and maintain a target network. Each agent stores experiences $e_t^i = (s_t, a_t^i, r_t^i, s_{t+1})$ in a replay buffer $D_i$. Mini-batches are randomly sampled from $D_i$ to train the Q-network, breaking correlations between sequential data and smoothing over changes in the data distribution. A separate network, the target network, with parameters $\theta_i^-$ that is periodically updated with the weights of the Q-network $\theta_i$. This helps in stabilizing the training by providing a consistent target during temporal difference updates.

In terms of the integration with the upper-level model, certain parameters in the lower level model are influenced by the upper level Market Mechanism Optimization (MMO) Model, creating a bilevel framework, including the penalties for deviations in renewable generation outputs, bidding price cap and the market clearing mechanism between pay-as-bid and pay-as-clear.

*B. Upper Level: Market Mechanism Optimisation*

The upper-level model, referred to as the Market Mechanism Optimization (MMO) model, is specifically designed to optimize the electricity market mechanisms that guide the simulation of the lower-level model. It focuses on day-ahead market operations within a given month, including GENCOs' bidding and market clearing processes. The MMO model targets three key aspects of market design: price cap on generator bids, settlement rules (pay-as-bid or pay-as-clear), and penalties for deviations in renewable energy output, all integral to promoting a sustainable and efficient energy market.

The optimization problem can also be formulated as a Markov Decision Process (MDP) defined by the tuple $\{S, A, R, P\}$. State space ($S$) represents the set of all possible market states. In our context, a state $s_T \in S$ at time $T$ includes market indicators aggregated from the lower-level model's simulation over a horizon (a month, 30 days), containing average market power indicator (Herfindahl-Hirschman Index, HHI), renewable energy penetration rates and average supply-demand ratios. Action space ($A$) is a set of all possible market mechanisms that can be implemented. An action $a_T \in A$ consists of the generator bid price cap ($PC_T$) indicating the maximum allowable bid price for participating GENCOs, settlement rule for market clearing ($MR_T$) between pay-as-bid ($MR_T = 0$) or pay-as-clear ($MR_T = 1$) and penalty for renewable deviations ($P_T$) applied to green GENCOs, which is formed as $a_T = \{PC_T, MR_T, P_T\}$. A scalar reward $r_T = R(s_T, a_T)$ is received after implementing action $a_T$ in state $s_T$. The reward is designed to reflect the overall day-ahead market performance within one month under the selected mechanisms, incorporating factors shown in (6). The probability $P(s_{T+1}|s_T, a_T)$ of transitioning to state $s_{T+1}$ from state $s_T$ after action $a_T$. This captures the dynamics of the market in response to the implemented mechanisms.

$$r_T = w_1 \cdot SW_T + w_2 \cdot RP_T \quad (6)$$

where $SW_T$ and $RP_T$ are social welfare and renewable penetration rate at time $T$; $w_1$ and $w_2$ are weights factors to balancing the importance of each component.

To achieve optimization, the MMO model employs a RL approach using the Proximal Policy Optimization (PPO) algorithm, a policy gradient method that utilizes an actor-critic framework. PPO is well-suited for environments with continuous and discrete action spaces, making it ideal for our application where actions include continuous variables (e.g., price cap $PC_T$, penalty $P_T$) and discrete choices (e.g., settlement rule $MR_T$).

PPO is a policy gradient method that enables efficient training of policies in environments with continuous action spaces, making it suitable for optimizing market mechanisms. The PPO algorithm aims to maximize the expected cumulative reward by updating the policy ($\pi_\theta(a_T|s_T)$) and value ($V_\phi(s_T)$) function parameters using the following clipped surrogate policy objective:

$$J^{actor}(\theta) = \mathbb{E}_T[\min(r_T(\theta, \theta_{old})\hat{A}_T, clip(r_T(\theta, \theta_{old}), 1 - \epsilon, 1 + \epsilon)\hat{A}_T] \quad (7)$$

where $r_T(\theta)$ is the probability ratio between the new and old policies, $\hat{A}_T$ is the estimated advantage at time $T$, $\in$ and $\theta_{old}$ refer to a small hyperparameter controlling the clipping range and the policy parameters before the update.

The value function (critic) is updated by minimizing the squared difference between the predicted value and the target value:

$$J^{critic}(\phi) = \mathbb{E}_T\left[\left(V_\phi(s_T) - V_T^{target}\right)^2\right] \quad (8)$$

$$V_T^{target} = r_T + \gamma V_{\phi_{old}}(s_{T+1}) \quad (9)$$

where $\gamma$ is the discount factor from 0 to 1, $V_{\phi_{old}}(s_{T+1})$ is the estimated value of the next state using the old value function parameters $\phi_{old}$.

An entropy term is added to encourage exploration by penalizing certainty in action selection:

$$J^{Entropy}(\phi) = \mathbb{E}_T[-\beta S[\pi_\theta](s_T)] \quad (10)$$

where $S[\pi_\theta](s_T)$ is the entropy of the policy at state $s_T$ and $\beta$ is a coefficient controlling the strength of the entropy bonus.

The combined loss function for optimization is:

$$\mathcal{L}(\theta,\phi) = -J^{actor}(\theta) + c_1 J^{critic}(\phi) + c_2 J^{Entropy}(\phi) \quad (11)$$

where $c_1$ and $c_2$ are coefficients balancing the contributions of the critic loss and entropy bonus.

The advantage function $\hat{A}_T$ measures the relative value of action $a_T$ in state $s_T$ compared to the expected value:

$$\hat{A}_T = Q(s_T, a_T) - V_\phi(s_T) \quad (12)$$

In terms of the policy and value function updates, the parameters $\theta$ and $\phi$ are updated using stochastic gradient ascent and descent, respectively:

$$\theta \leftarrow \theta + \alpha_\theta \nabla_\theta J^{actor}(\theta) \quad (13)$$

$$\phi \leftarrow \phi - \phi_\alpha \nabla_\phi J^{critic}(\phi) \quad (14)$$

where $\alpha_\theta$ and $\phi_\alpha$ are the learning rate for the actor and critic.

## III. Quantum Computing Algorithm Design

To further enhance the lower-level model's capability in capturing the complex dynamics of the electricity market, we integrate quantum computing techniques into the Deep Q-Network (DQN) framework. Specifically, we employ a Variational Quantum Circuit (VQC) with six qubits—each representing a GENCO—to model the agents' decision-making processes.

In the quantum setting, the Markov Decision Process (MDP) elements—states, actions, and rewards—are encoded into quantum states using qubits. The Quantum MDP is defined as a tuple $\{\dot{S}, \dot{A}, \dot{R}, \dot{P}\}$, mirroring the classical MDP but within a quantum framework. Each classical state $s \in S$ is mapped to a quantum state $|\psi_s\rangle \in \mathcal{H}$ denotes the Hibert space of the qubits. Actions remain classical decisions but are used to manipulate quantum states via quantum gates. Rewards are obtained from measurements of the quantum states after applying certain operations corresponding to the actions taken. Transition probabilities are inherently probabilistic due to the quantum nature of state evolution.

The VQC serves as a quantum function approximator for the Q-function $\mathbb{Q}(\dot{s}, \dot{a}; \dot{\theta})$, where $\dot{\theta}$ represents the set of variational parameters (quantum gate parameters) in the circuit. The operation of VQC for enabling quantum properties contain three main steps: encoding, entanglement and observation.

Classical input data $s$ is pre-processed and encoded into quantum states. Each element of $s$ is normalized to the range $[0, \pi]$ to obtain $s_\omega$. The encoding is performed using rotation gate applied to each qubit:

$$|\psi_{enc}\rangle = \otimes_{i=1}^{6} R_x\left(s_\omega^{(i)}\right)|0\rangle_i \quad (15)$$

where $R_x$ is the rotation gate about the x-axis:

$$R_x(s_\omega) = \begin{bmatrix} \cos\left(\frac{s_\omega}{2}\right) & -i\sin\left(\frac{s_\omega}{2}\right) \\ -i\sin\left(\frac{s_\omega}{2}\right) & \cos\left(\frac{s_\omega}{2}\right) \end{bmatrix} \quad (16)$$

After encoding, the qubits enter the variational quantum circuit where they undergo further rotations and entanglement operations. We introduce both $R_y$ and $R_z$ gates, parameterized by $\omega$, to enhance the expressiveness of the circuit. The unitary operation $U(\omega)$ of the VQC is defined as:

$$U(\omega) = \prod_{l=1}^{L}\left(\left[\otimes_{i=1}^{6} R_y\left(\omega_i^{(l)}\right) R_z\left(\omega_i^{(l)}\right) \cdot \varepsilon\right]\right) \quad (17)$$

$$R_y(\omega) = \begin{bmatrix} \cos(\omega/2) & -\sin(\omega/2) \\ \sin(\omega/2) & \cos(\omega/2) \end{bmatrix} \quad (18)$$

$$R_z(\omega) = \begin{bmatrix} e^{-i\omega/2} & 0 \\ 0 & e^{i\omega/2} \end{bmatrix} \quad (19)$$

$$\varepsilon = \prod_{i=1}^{5} CNOT_{i,i+1} \quad (20)$$

where $L$ is the number of layers in the circuit, $\omega_i^{(l)}$ is the variational parameter (rotation angle) for qubit $i$ in layer $l$, $R_y$ and $R_z$ are rotation gates about the y-axis and z-axis, respectively; $\varepsilon$ defines the entanglement operation, implemented via Controlled-NOT (CNOT) gates arranged in a linear (ladder) topology.

In each layer $l$, the qubits are individually rotated by $R_y$ and $R_z$ gates with parameters $\omega_i^{(l)}$, introducing non-linearity and complexity into the quantum state. The subsequent entanglement operation $\varepsilon$ creates correlations between qubits, enabling the circuit to capture complex patterns and interactions relevant to the bidding strategies of the GENCOs.

After the application of $U(\omega)$, measurements are performed on the final quantum state $|\psi_{\dot{s}}\rangle$ to extract

expectation values corresponding to the Q-values. The observable associated with action $\dot{a}$ is:

$$O_a = \sum_{i=1}^{6} w_o^{(\dot{a},i)} \sigma_z^{(i)} \qquad (21)$$

$$\sigma_z = \begin{bmatrix} 1 & 0 \\ 0 & -1 \end{bmatrix} \qquad (22)$$

where $\sigma_z^{(i)}$ is the Pauli-Z operator acting on qubit $i$ and $\omega_o^{(\dot{a},i)}$ are trainable weights scaling the contribution of each qubit for action $\dot{a}$.

The final Q-value at time $t$ for agent $i$ is computed as:

$$\mathbb{Q}(\dot{s}_t, \dot{a}_t^i; \dot{\theta}_i, w_{o_i}) = \langle O_{\dot{a}_t}^i \rangle \qquad (23)$$

The corresponding loss function for training the quantum-enhanced DQN is defined in (24) and the parameter update rules can be implemented by (25) and (26).

$$L_i(\dot{\theta}_i, w_{o_i}) = \mathbb{E}_{(s_t, a_t^i, r_t^i, s_{t+1})}[(y_t^i - \langle O_{\dot{a}_t}^i \rangle)^2] \qquad (24)$$

$$\dot{\theta}_i \leftarrow \dot{\theta}_i - \alpha \nabla_{\dot{\theta}_i} L_i(\dot{\theta}_i, w_{o_i}) \qquad (25)$$

$$w_{o_i} \leftarrow w_{o_i} - \alpha \nabla_{w_{o_i}} L_i(\dot{\theta}_i, w_{o_i}) \qquad (26)$$

where $\alpha$ is the learning rate.

## IV. CASE STUDY

A test market study was developed based on the IEEE 30-bus system with 6 GENCO agents with specific generation costs in [6], for comparison of the proposed bilevel model with VQC installed in the lower level and the bilevel model using conventional reinforcement learning techniques. The hourly load demand profile is also extracted from [6]. For the initial configuration of the market environment used in model training, the settlement rule is selected as pay-as-bid. The bidding range is established within the interval [0, 100] USD/MWh, and the penalty coefficient for deviations in renewable energy output is set at 10% of the deviation. As the upper-level model progresses through the learning process, it dynamically adjusts the settlement rules between pay-as-bid and pay-as-clear. Additionally, the bidding cap is redesigned to fall within the range of [50, 500] USD/MWh, and the penalty coefficient for renewable energy output deviations is fine-tuned within the interval of [5, 15] %.

The output are the converged social welfare of the proposed three market mechanisms and the following early stopping criteria is to identify the convergence: the social welfare within the one month has a lower then 20% change over the three consecutive training episodes of the upper model. Table I records the final market design and the social welfare of the two models. The comparative results between the quantum-enhanced reinforcement learning model and the classical reinforcement learning approach in designing electricity market mechanisms reveal substantial differences in key parameters and outcomes. Specifically, the quantum-enhanced bilevel model employs a pay-as-clear settlement rule with a price cap range of [0, 396] USD/MWh, a renewable energy output deviation penalty of 9%, and achieves a social welfare of 3,520,736 USD. In contrast, the classical reinforcement learning model utilizes the same settlement rule but with a narrower price cap range of [0, 125] USD/MWh, a higher penalty of 15%, and results in a significantly lower social welfare of 1,354,578 USD. The quantum-enhanced model demonstrates a markedly broader price cap range compared to the classical approach. This expanded range allows GENCOs greater flexibility in their bidding strategies, facilitating a more comprehensive exploration of price dynamics within the market. The increased upper limit of 396 USD/MWh as opposed to 125 USD/MWh suggests that the quantum-enhanced model can accommodate a wider spectrum of bidding behaviours, potentially leading to more competitive and efficient market outcomes. The ability to handle a larger price cap range may reduce instances of price suppression and enhance the market's ability to reflect true supply and demand conditions. The quantum-enhanced model employs a 9% penalty for deviations in renewable energy output, which is lower than the 15% penalty utilized in the classical model. This reduction in penalty serves multiple purposes: encouraging renewable participation, promoting accurate forecasting and maintaining grid stability. Lower penalties mitigate the financial risks associated with the inherent variability of renewable energy sources, thereby fostering greater participation from renewable GENCOs. A more balanced penalty structure incentivizes renewable providers to improve forecasting accuracy without imposing excessively stringent financial burdens that could deter market participation. While penalties are essential for ensuring reliability, the optimized level of 9% strikes a balance between enforcing adherence to forecasts and allowing flexibility in energy production, thereby maintaining grid stability without discouraging renewable integration.

The most striking difference lies in the social welfare outcomes, which can be attributed to several factors including improved market efficiency, balanced mechanism design and advanced optimisation capabilities. The quantum-enhanced model's ability to explore a more extensive parameter space likely leads to more efficient market clearing, reducing inefficiencies and maximizing overall economic welfare. y optimizing both the price cap and penalty parameters, the quantum-enhanced model effectively balances the interests of GENCOs and consumers, enhancing social welfare through better alignment of incentives and market operations. Quantum computing's inherent ability to handle complex, high-dimensional optimization problems enables the model to identify optimal configurations of market mechanisms that classical reinforcement learning may overlook or approximate less effectively.

TABLE I COMPARISON OF TWO ALGORITHMS.

|  | w/ VQC | w/o VQC |
| --- | --- | --- |
| Settlement rule | Pay-as-clear | Pay-as-clear |
| Price cap | [0,396] USD/MWh | [0,125] USD/MWh |
| Penalty coefficient | 9% | 15% |
| Social welfare | 3520736 USD | 1354578 USD |

## V. CONCLUSION

This study presents a novel bilevel optimization framework tailored for electricity market operations, integrating advanced reinforcement learning techniques with quantum computing to enhance market mechanism design. The upper-level Market

MMO model employs PPO to dynamically adjust key market parameters, including settlement rules, generator bid price caps, and penalties for deviations in renewable energy output. Concurrently, the lower-level model leverages a VQC-enhanced DQN to accurately simulate and optimize the bidding strategies of GENCOs. The comparative analysis between quantum-enhanced and classical reinforcement learning approaches underscores the significant advantages of our bilevel model. Specifically, the quantum-enhanced model operates with a broader price cap range and a more balanced renewable output deviation penalty, resulting in a substantial increase in social welfare. In contrast, the classical model's narrower price cap and higher penalty yield a lower social welfare. These results highlight the superior parameter flexibility and market efficiency achieved through quantum-enhanced optimization, facilitating better integration of renewable energy sources and enhancing overall market stability.

The inherent capabilities of quantum computing, such as superior function approximation and enhanced exploration through quantum superposition and entanglement, enable the bilevel model to navigate complex, high-dimensional optimization landscapes more effectively than traditional methods. This results in more precise policy optimization and significant improvements in social welfare, demonstrating the potential of quantum-enhanced reinforcement learning in revolutionizing energy market design. Moreover, the bilevel structure facilitates a responsive and adaptive market mechanism, where the upper-level model continuously refines market parameters based on real-time feedback from the lower-level simulations. This dynamic interplay ensures that the market mechanisms remain robust and efficient amidst evolving market conditions and increasing renewable penetration.